\documentclass[%
 reprint,
 amsmath,amssymb,
 aps,
]{revtex4-1}

\usepackage{graphicx}
\usepackage{dcolumn}
\usepackage{bm}

\usepackage{mathptmx}
\usepackage{amsmath,amsfonts,amsthm, amssymb,latexsym}

\def\be{\begin{equation}}
\def\ee{\end{equation}}

\begin{document}

\preprint{APS/123-QED}

\title{The internal composition of proto-neutron stars under strong magnetic fields }

\author{B. Franzon}
\affiliation{ Frankfurt Institute for Advanced Studies,
Ruth-Moufang - 1 60438, Frankfurt am Main,
Germany}
\email{franzon@fias.uni-frankfurt.de}

\author{V. Dexheimer}
 \affiliation{Department of Physics, Kent State University, Kent OH 44242 USA}
\email{vdexheim@kent.edu}

\author{ S. Schramm }%
\affiliation{%
 Frankfurt Institute for Advanced Studies,
Ruth-Moufang - 1 60438, Frankfurt am Main,
Germany
}%
 \email{schramm@fias.uni-frankfurt.de}

%

%




\date{\today}

\begin{abstract}
In this work, we study the effects of magnetic fields and rotation  on the structure and composition of proto-neutron stars (PNS's). A hadronic chiral SU(3) model is applied to cold neutron stars (NS) and proto-neutron stars with trapped neutrinos and at fixed entropy per baryon.  We obtain general relativistic solutions for neutron and proto-neutron stars endowed with a poloidal magnetic field by solving Einstein-Maxwell field equations in a self-consistent way.  As the neutrino chemical potential decreases in  value over time, this alters the chemical equilibrium and the composition inside the star, leading to a change in the structure and in the particle population of these objects. We find that the magnetic field deforms the star and significantly alters the number of trapped neutrinos in the stellar interior, together with strangeness content and temperature in each evolution stage. 
\end{abstract}

\maketitle


\section{Introduction}
Proto-neutron stars (PNS) are newborn compact stars generated immediately after the gravitational collapse of the core of  massive stars, which cool down and contract to become  neutron stars (NS). On a time scale of 10-20 seconds, PNS's cool significantly and lose their high lepton content mainly through electron neutrino ($\nu$) emission  \cite{burrows1986birth, pons1999evolution}.  The entropy per baryon in  PNS's is of the order of 1 or 2, making them, therefore, very hot stars (T up to 50 MeV in the center). The environment in these stars is so extreme, that neutrinos can be trapped on dynamical time scales and develop a finite chemical potential \cite{prakash1997composition}. In addition, it has been shown that rotation can play an important role in the description of these objects \cite{Goussard:1996dp, Goussard:1997bn}.

Although the initial evolution of PNS's from  hot, $\nu$-trapped  and lepton-rich to cold and $\nu$-free NS's is far from equilibrium and characterized by strong instabilities, just a few seconds after the bounce, they can be approximately considered as a sequence of equilibrium configurations.  This is the so-called Kelvin-Helmholtz phase \cite{pons1999evolution, fischer2010protoneutron}. During this process, the structure of the PNS can be divided into a core region, that will be studied in this work, and an envelope with entropy per baryon much higher than in the core.  In the core, the entropy per baryon can reach values of $s_{B}\backsimeq 1,2$. A fixed entropy per baryon allows to model a temperature increase towards the center of the star. These properties make PNS's quite different objects from the ordinary neutron stars, which are usually observed as radio pulsars. However, as NSs are born from PNS's, one expects that some features currently presented in neutron stars as, for example, magnetic fields and rotation rates,  are related to their progenitors. 


It is generally believed that certain classes of neutron stars possess very strong magnetic fields on their surfaces on the other of $10^{12-15}$ G \cite{shapiro2008black}. Such fields are usually estimated from observations of the stars' period and  period derivative. However, the internal magnetic field in these stars can be even stronger. For example, according to the virial theorem, they can have central magnetic fields of the order of $10^{18}$ G  \cite{ferrer2010equation, lai1991cold, fushiki1989surface, cardall2001effects}. 

According to Ref.~\cite{woltjer}, such strong magnetic fields originate from the conservation of magnetic flux during the collapse of the core of a supernova. But this idea is not suitable for highly magnetized neutron stars, since a surface magnetic field of the order of $\mathrm{10^{15}}$ G would require a radius less than the Schwarzschild radius for a canonical neutron star with $\rm{M \sim 1.4\, M_{\odot}}$.  Another idea was suggested by Thompson and Duncan \cite{Duncan:1992hi}, in which a proto-neutron star can combine convection and differential rotation in order to generate a dynamo process, which is able to produce fields as large as $\mathrm{10^{15}}$ G. However, as shown in Ref.~\cite{vink2006supernova}, this explanation does not explain the supernova remnants associated with these objects. 

Recently, it was shown that magnetorotational instabilities (MRI) in proto-neutron stars can amplify small seed magnetic fields over very short time scales \cite{akiyama2003magnetorotational, obergaulinger2009semi, sawai2013global, sawai2016evolution, cerda2008new, mosta2015large}. However,  the limit of this amplification is still unknown. As stated in Ref.~\citep{rembiasz2016maximum},  the amplification factor seems to be   small and, therefore, the magnetic field cannot be amplified through MRI channel modes. In this case, the authors in Ref.~\citep{rembiasz2016maximum} suggest that another physical process, as a MRI-driven turbulent dynamo, could further amplify small seed magnetic fields in PNS's.

In Ref.~\cite{franzon2016self}, we studied the effects of strong magnetic fields on hybrid stars by using a full  general-relativity approach,  solving the coupled Maxwell-Einstein equation in a self-consistent way. The magnetic field was assumed to be axi-symmetric and poloidal. We took into consideration the anisotropy of the energy-momentum tensor due to the magnetic field, magnetic field effects on the equation of state, the interaction between matter and the magnetic field (magnetization), and the anomalous magnetic moment of the  hadrons. The equation of state used was an extended hadronic and quark chiral SU(3) non-linear realization of the sigma model that describes hybrid stars containing nucleons, hyperons and quarks (see Refs.~\cite{PhysRevC.88.014906,Papazoglou:1998vr,Dexheimer:2008ax,Dexheimer:2009hi}). According to our results, the effects of the magnetization and  the magnetic field on the EoS do not play an important role for global properties of these stars.  On the other hand,  the magnetic field causes the central density in these objects to be reduced, inducing major changes in the populated degrees of freedom and, potentially, converting a hybrid star with hadronic and quark phases into a hadronic star. 

The composition and the structure of PNSs are strongly related to the number of trapped neutrinos.  As the neutrino chemical potential decreases over time, this alters the chemical equilibrium, leading to an impact on the structure and on the composition of these stars. In this context, as we showed in Ref.~\cite{franzon2016self}, strong magnetic fields have a huge impact not only on the structure of NS's, but also on the particle population inside cold NS's. In this work, we study the effects of strong magnetic fields on a hot and rapidly rotating proto-neutron star, since the magnetic field can affect the amount of trapped neutrinos and prevent or favour exotic phases with hyperons or quarks. For this purpose, we make use of the hadronic  chiral SU(3) model \cite{PhysRevC.88.014906,Papazoglou:1998vr,Dexheimer:2008ax,Dexheimer:2009hi} explicitly including  trapped neutrinos and fixed entropy per baryon. The cold and hot EoS's  are then calculated at finite temperature and over a range of entropies and neutrino fractions. Finally, we construct proto-neutron stars models by using the LORENE C++ library, which solves numerically the Einstein-Maxwell equations by means of a pseudo-spectral method as in Refs.~\cite{Bonazzola:1993zz, Bocquet:1995je}. Recently, we applied this approach to magnetized hibrid stars in Ref.~\cite{franzon2016self} and to magnetized and fast rotating white dwarfs in Ref.~\cite{Franzon:2015gda}.

In Ref.~\cite{pons2001evolution} the authors addressed the importance of quarks in the evolution process of PNS's. The appearance of quarks softens the equation of state and may lead to less massive and smaller stars \cite{lattimer2001neutron}.  In addition, quarks would alter the neutrino emissivities and, therefore, influence other properties like the surface temperature in PNS's and NS's. In a future work, we will investigate the role played by phase  transitions from quark to hadronic matter inside the stars, but in this work, we neglect possible effects of a quark phase. However, in our hadronic model we include hyperons as the "exotic matter" component that can, potentially, soften the EoS. 
 Note that, there is no reason to ignore the appearance of hyperons, as they should appear at about two times saturation density, and their presence might produce distinct neutrinos signals that can detected in the next generation neutrino detectors \cite{diwan2000next}. 

The article  is organised as follows. In Sec. II we present the equation of state and the model Lagrangian used in the work. In section III we briefly discuss how to solve the Einstein-Maxwell field equations. In Sec. IV, we report our results and
discuss their consequences. Finally in Sec. V, we summarize our findings and present conclusions.

\section{Stellar interior: equation of state}
Chiral sigma models are effective relativistic models that describe hadrons interacting via meson exchange and, most importantly, are constructed from symmetry relations. They are constructed in a chirally invariant manner since the particle masses originate from interactions with the medium and, therefore, go to zero at high density and/or temperature. 

The non-linear realization of the sigma model is an improvement over the widely-used sigma model and it includes the pseudoscalar mesons as the angular parameters for the chiral transformation. As a result, these mesons only appear if the symmetry is broken or in terms of derivatives of the fields and the scalar and pseudoscalar sectors decouple from each other, leading to a greater freedom in the manner in which baryons and mesons couple to each other. As a consequence of those couplings, the non-linear realization of the sigma model is in very good agreement with nuclear physics data, such as the vacuum masses of the baryons, saturation properties, hyperon potentials, pion and kaon decay constants $f_{\pi}$ and $f_{k}$, etc \cite{Papazoglou:1998vr,Bonanno:2008tt}.

The Lagrangian density of the SU(3) non-linear realization of the sigma model in the mean field approximation, applied to neutron star matter can be found in Refs.~\cite{Dexheimer:2008ax,Schurhoff:2010ph,Dexheimer:2015qha}. A recent extension of this model also includes quarks as dynamical degrees of freedom \cite{Dexheimer:2009hi, PhysRevC.88.014906, Negreiros:2010hk,Steinheimer:2010ib,PhysRevC.88.014906,Dexheimer:2014pea}. In this work, we make use of the simple hadronic version of the model, as it was shown in Ref.~\cite{franzon2016self} that strong magnetic fields strongly supress deconfinement to quark matter in neutron stars. The  Lagrangian density of the model we use in this work reads:
\begin{eqnarray}
\mathcal{L} = \mathcal{L}_{\rm{Kin}}+\mathcal{L}_{\rm{Int}}+\mathcal{L}_{\rm{Self}}+\mathcal{L}_{\rm{SB}}\,,
\end{eqnarray}
where, besides the kinetic energy term for hadrons and leptons (included to ensure charge neutrality), the terms:
\begin{eqnarray}
\mathcal{L}_{\rm{Int}}&=&-\sum_i \bar{\psi_i}[\gamma_0(g_{i\omega}\omega+g_{i\phi}\phi+g_{i\rho}\tau_3\rho)+M_i^*]\psi_i,\nonumber\\
\mathcal{L}_{\rm{Self}}&=&+\frac{1}{2}(m_\omega^2\omega^2+m_\rho^2\rho^2+m_\phi^2\phi^2)\nonumber\\
&-&k_0(\sigma^2+\zeta^2+\delta^2)-k_1(\sigma^2+\zeta^2+\delta^2)^2\nonumber\\&-&k_2\left(\frac{\sigma^4}{2}+\frac{\delta^4}{2}
+3\sigma^2\delta^2+\zeta^4\right)
-k_3(\sigma^2-\delta^2)\zeta\nonumber\\&-&k_4\ \ \ln{\frac{(\sigma^2-\delta^2)\zeta}{\sigma_0^2\zeta_0}}\nonumber\\&+&g_4 (\omega^4 + 3 \omega^2 \phi^2 + \phi^4/4 + 4 \omega^3 \phi/\sqrt{2} + 2 \omega \phi^3/\sqrt{2}),\nonumber\\
\mathcal{L}_{\rm{SB}}&=&-m_\pi^2 f_\pi\sigma-\left(\sqrt{2}m_k^ 2f_k-\frac{1}{\sqrt{2}}m_\pi^ 2 f_\pi\right)\zeta\,,
\end{eqnarray}
represent the interactions between baryons and vector and scalar mesons, the self interactions of scalar and vector mesons, and an explicit chiral symmetry breaking term, which is responsible for producing the masses of the pseudo-scalar mesons. The index $i$ denotes the states of the baryon octet. The electrons and muons are included as a free Fermi gas. The meson fields included are the vector-isoscalars $\omega$ and $\phi$ (strange quark-antiquark state), the vector-isovector $\rho$, the scalar-isoscalars $\sigma$ and $\zeta$ (strange quark-antiquark state) and  the scalar-isovector $\delta$, with $\tau_3$ being twice the isospin projection of each particle. The isovector mesons affect isospin-asymmetric matter and, thus, are important for neutron star physics. Also, the $\delta$ meson has a contrary but complementary role to the $\rho$ meson, much like the $\sigma$ and $\omega$  mesons.

The effective masses of the baryons are simply generated by the scalar mesons, except for a small explicit mass term $M_0$: 
\begin{eqnarray}
M_{i}^*&=&g_{i\sigma}\sigma+g_{i\delta}\tau_3\delta+g_{i\zeta}\zeta+M_{0_i}\,.\end{eqnarray}
The scalar sector of the coupling constants ($g_{N \sigma}=-9.83$, $gN_{\delta}=-2.34$, $gN_{\zeta}=1.22$, $k_0=1.19 \chi^2$, $k_1=-1.40$, $ k_2=5.55$, $k_3=2.65 \chi$ and $k_4=-0.06 \chi^4$, with $\chi = 401.93$ MeV, $M_{0}=150$ and 354 MeV for nucleons and hyperons, respectively.) is connected through  SU(3) symmetry and determined to reproduce the vacuum masses of the baryons and scalar mesons, and the pion and kaon decay constants $f_\pi$ and $f_\kappa$. The vector sector of the model ($g_{N \omega}=11.90$, $g_{N \rho}=4.03$, $g_{N \phi}=0$ and $g_4=38.90$) is connected mainly through SU(6) symmetry and determined to reproduce nuclear saturation properties ($\rho_0=0.15$ fm$^{-3}$, $B/A=-16.00$ MeV, $K=297.32$ MeV, $E_{\rm{sym}}=32.5$ MeV, $L=93.85$ MeV) and astrophysical observations.  We also reproduce reasonable values for the hyperon potentials $U_\Lambda=-28$ MeV, $U_\Sigma=5.35$ MeV and $U_\Xi=-18.36$ MeV, which are are calculated as $U_i=M^*_i+g_{i\omega}\omega+g_{i\phi}\phi-{M_0}_i$ for symmetric matter at
saturation.

In order to obtain values for the mesonic fields at a certain temperature and baryon chemical potential, we solve a system of coupled equations, including the equations of motion for the mesonic fields \cite{Dexheimer:2008ax}. We further impose charge neutrality $\Sigma_i Q_{e i} n_i=0$ (with $Q_{e}$ being the electric charge and $n_i$ the number density of each species) and chemical equilibrium on the system. Finite-temperature calculations include the heat bath of hadronic quasiparticles within the grand canonical potential of the system. To simulate proto-neutron star conditions, we include trapped neutrinos by fixing the lepton fraction $Y_l=\Sigma_i Q_{l i} n_{i} / n_B$ (with the lepton number $Q_{l }$ being non-zero only for leptons) \cite{Burrows:1986me,Keil:1995hw, gudmundsson1980consequence}. We also fix the entropy per baryon $s_{B}=S/A=s/n_B$ in the core of the star \cite{Burrows:1986me,pons2001evolution,pons1999evolution,Stein:2005nt}. 

\begin{figure}
\begin{center}\includegraphics[width=0.7\textwidth,angle=-90,scale=0.5]{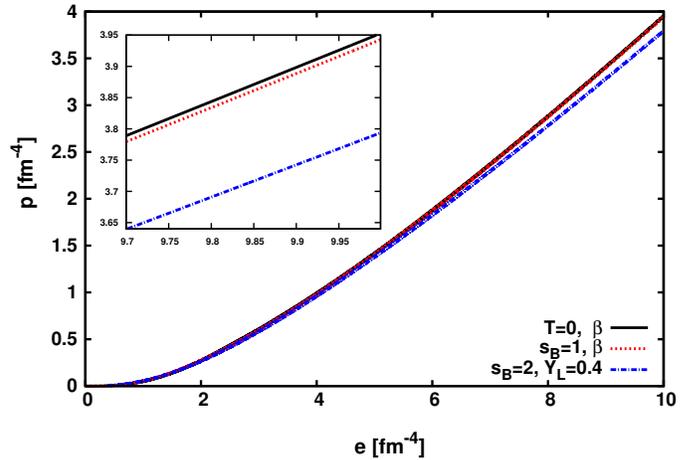}
\caption{\label{eos} Equations of state for proto-neutron and neutron stars. Note that the $T=0$ and $s_{B}=1$ in $\beta$-equilibrium lines almost overlap.}\end{center}
\end {figure}

The temperature is not expected to be constant in the interior of compact stars. Sophisticated approaches have realistic profiles for temperature \cite {Reddy:1997yr, Pons:2000xf} but, in this work, we do not attempt to make use of them since our aim is only to investigate magnetic field effects on different approximate stages of the star evolution. For this reason, as an approximation, we are going to consider different values of fixed entropy per baryon throughout the star. In Fig.~\ref{eos} we show the three equations of state used in this work. 

\section{General Relativistic Calculation} 
The formalism used in this work  was first applied to rotating and non-rotating magnetized neutron stars in Refs.~\cite{Bonazzola:1993zz, Bocquet:1995je, Chatterjee:2014qsa}, and more recently in Ref.~\cite{franzon2016self}. It allows us to obtain equilibrium configurations by solving the Einstein-Maxwell field equations for  spherical  polar  coordinates with the origin  at  the  stellar  center and with the  pole  along  the  axis  of  symmetry. For more details on the gravitational equations and numerical procedure, see Ref.~\cite{gourgoulhon20123+}.  Here, we present the basic electromagnetic equations that, together with the gravitational equations, are required to be solved numerically. In this context, the stress-energy tensor $T_{\alpha\beta}$ contains both the matter (without the magnetic field effects) and electromagnetic source terms:
\be
T_{\alpha\beta} = (e+p)u_{\alpha}u_{\beta} + pg_{\alpha\beta} + \frac{1}{\mu_{0}} \left( F_{\alpha \mu} F^{\mu}_{\beta} - \frac{1}{4} F_{\mu\nu} F^{\mu\nu} \mathrm{g}_{\alpha\beta} \right),
\label{emt}
\ee
with  $F_{\alpha\mu}$ being the antisymmetric Faraday tensor defined as $F_{\alpha\mu} = \partial_{\alpha} A_{\mu} - \partial_{\mu} A_{\alpha} $, where $A_{\mu}$ is the electromagnetic four-potential. As we are dealing with stars endowed with poloidal magnetic fields, one has $A_{t}$ and $A_{\phi}$ as the two non-zero components of the electromagnetic four-potential, $A_{\mu} = (A_{t}, 0 , 0 , A_{\phi})$. The total energy density of the system is $e$, the isotropic contribution to the pressure is denoted by $p$,  $u_{\alpha}$ is the fluid 4-velocity, and the metric tensor is $g_{\alpha\beta}$. The first term in Eq.~\eqref{emt}  represents the  isotropic matter contribution to the energy momentum-tensor,  while the second term is the anisotropic electromagnetic field contribution. Note,  that we are not including anisotropies due to the magnetization as done in Refs.~\cite{Chatterjee:2014qsa, franzon2016self}.  This is due to the fact that in this work we are not taking into account magnetic field effects in the EoS. In  Ref.~\cite{Chatterjee:2014qsa} and later in Ref.~\cite{franzon2016self}, it was already shown that there is none or a small contribution when taking into account the magnetic field corrections in the equation of state through the magnetization. The metric tensor in this axi-symmetric spherical-like coordinates $(r, \theta, \phi)$ can be expressed as:
\begin{align}
ds^{2} = &-N^{2} dt^{2} + \Psi^{2} r^{2} \sin^{2}\theta (d\phi - N^{\phi}dt)^{2} \\
 &+ \lambda^{2}(dr^{2} + r^{2}d\theta^{2}), \nonumber
\label{line}
\end{align}
with N, $N^{\phi}$, $\Psi$ and $\lambda$ being functions of the coordinates $(r, \theta)$. 

According to \cite{lichnerowicz1967relativistic}, the electric field components can be written as:
\be
E_{\alpha} = \left( 0 , \frac{1}{N} \left[  \frac{\partial A_{t}}{\partial r} + N^{\phi} \frac{\partial A_{\phi}}{\partial r}\right ] , \frac{1}{N} \left[  \frac{\partial A_{t}}{\partial \theta} + N^{\phi} \frac{\partial A_{\phi}}{\partial \theta}\right ]   , 0 \right),
\ee
and the magnetic field reads:
\be
\hspace{-2cm} B_{\alpha} = \left( 0 , \frac{1}{\Psi r^{2} \sin \theta} \frac{\partial A_{\phi}}{\partial \theta}, - \frac{1}{\Psi \sin \theta} \frac{\partial A_{\phi}}{\partial r} , 0  \right).
\ee

The  Faraday tensor $F_{\alpha\mu}$ can be derived from the electromagnetic four-potential $F_{\alpha\mu} = A_{\alpha, \mu} - A_{\mu,\alpha}$, so that the homogeneous Maxwell equation:
\be
F_{\alpha\mu;\gamma} + F_{\mu\gamma;\alpha} + F_{\gamma\alpha;\mu} = 0,
\ee
is automatically satisfied. According to Ref.~\cite{Bocquet:1995je}, the inhomogeneous
Maxwell equation:
\be 
\nabla_{\mu}  F^{\alpha\mu} = \mu_{0}j^{\alpha},
\ee
can be expressed in terms of the two non-vanishing components of the electromagnetic potential $A_{\mu}$ through the Maxwell-Gauss equation:
\begin{align}
\Delta_{3} A_{t} = &-\mu_{0}\lambda^{2}(j_{t}+j_{\phi})-\frac{\Psi^{2}}{N^{2}}N^{\phi}r^{2}\rm{sin}^2\theta \partial A_{t} \partial N^{\phi} \nonumber \\
&-\left( 1+ \frac{\Psi^{2}}{N^{2}}r^{2}\rm{sin}^2\theta (N^{\phi})^{2}  \right)\partial A_{\phi}\partial N^{\phi} \nonumber \\
&-(\partial A_{t} + 2 N^{\phi} \partial A_{\phi})\partial(\beta-\nu) \nonumber \\
&-2\frac{N^{\phi}}{r} \left(\frac{\partial A_{\phi}}{\partial r}  + \frac{1}{\rm{tan}\theta}\frac{\partial A_{\phi}}{\partial \theta} \right ),
\label{maxwellgauss}
\end{align}
and through the Maxwell-Amp\`ere equation as:
\begin{align}
\tilde{\Delta_{3}} \left( \frac{A_{\phi}}{r \rm{sin}\theta} \right) =&-\mu_{0}\lambda^{2}\Psi^{2}(j^{\phi}-N^{\phi}j^{t})r \rm{sin}\theta \nonumber\\
& + \frac{\Psi^{2}}{N^{2}}r \rm{sin}\theta \partial N^{\phi} (\partial A_{t}+N^{\phi}\partial A_{\phi}) \nonumber \\
& + \frac{1}{r \rm{sin}\theta} \partial A_{\phi} \partial (\beta - \nu),
\label{maxwellampere}
\end{align}
with the notation:
\begin{align}
&\hspace{-2cm}\Delta_{3}=\frac{\partial^{2}}{\partial r^{2}} + \frac{2}{r}\frac{\partial}{\partial r}  + \frac{1}{r^{2}}\frac{\partial^2}{\partial \theta^{2}}  + \frac{1}{r^{2} \rm{tan}\theta}\frac{\partial}{\partial \theta}, \nonumber \\
&\hspace{-2cm}\tilde{\Delta_{3}}=\Delta_{3} -\frac{1}{r^{2}\rm{sin}^{2}\theta}, \nonumber \\
&\hspace{-2cm}\nu=\rm{ln}A, \beta=\rm{ln}\Psi, \alpha=\rm{ln} \lambda,  \nonumber \\
&\hspace{-2cm}\Delta_{2}=\frac{\partial^{2}}{\partial r^{2}} + \frac{1}{r}\frac{\partial}{\partial r} + \frac{1}{r^{2}}\frac{\partial^{2}}{\partial r^{2}}, \nonumber \\
&\hspace{-2cm} \partial a \partial b = \frac{\partial a}{\partial r} \frac{\partial b}{\partial r} + \frac{1}{r^{2}}\frac{\partial a}{\partial \theta} \frac{\partial b}{\partial \theta}. \nonumber
\end{align}
Within the $3+1$ decomposition and under the assumptions of  stationary and axisymmetric space-time, the Einstein equations for the metric potentials in Eq.5 are given by \cite{Bonazzola:1993zz, gourgoulhon20123+, Chatterjee:2014qsa}:
\begin{align}
&\Delta_{3}\nu=4\pi G \lambda^{2} (E+S^{i}_{i}) + \frac{\Psi^2 r^{2} \rm{sin}^2\theta}{2N^{2}}(\partial N^{\phi})^{2} - \partial\nu \partial(\nu+\beta), \\
& \tilde{\Delta_{3}}(N^{\phi}r\rm{sin}\theta)= -16\pi G\frac{N\lambda^2}{\Psi}\frac{J_{\phi}}{r\rm{sin}\theta}- r \rm{sin}\theta \partial N^{\phi} \partial(3\beta-\nu), \\
& \Delta_{2}[(N \Psi-1) r \rm{sin} \theta]=8\pi G N \lambda^{2} \Psi r \rm{sin}\theta (S^{r}_{r}+ S^{\theta}_{\theta}), \\
&  \Delta_{2}(\nu+\alpha)=8\pi G \lambda^{2} S^{\phi}_{\phi} + \frac{3\Psi^{2} r^2 \rm{sin}^2 \theta}{4N^{2}} (\partial N^{\phi})^{2} - (\partial\nu)^{2},
\end{align}
being $E$ total energy density of the fluid given by:
\be
E= \Gamma^2 (e + p) - p,
\label{energydensitypf}
\ee
while the momentum density flux is:
\be 
J_{\phi} = \Gamma^2 (e + p) U .
\label{momentumdensitypf}
\ee
The  3-tensor stress components are expressed as:
\be
S^{r}_{r} = S^{\theta}_{\theta}  = p ,
\ee
and 
\be
S^{\phi}_{\phi} = p + (E + p) U^2,
 \ee
where the Lorentz factor is given by $\Gamma = (1 - U^2)^{-\frac{1}{2}}$ being $U$ the fluid velocity  defined  as:
\be
U = \frac{\Psi r\sin\theta}{N}(\Omega - N^\phi),
\ee
with the lapse function $N^\phi$ and  the angular velocity $\Omega$, as measured by an observer at infinity (see Refs.~\cite{cardall2001effects,Bonazzola:1993zz, gourgoulhon20123+} for more details). As in Ref.~\cite{Bonazzola:1993zz},  the equation of motion ($\nabla_{\mu}T^{\mu\nu}= 0$) reads:
\be
H \left(r, \theta \right) + \nu \left(r, \theta \right) -\ln \Gamma \left( r, \theta \right) + M \left(r, \theta \right) = const .
\label{equationofmotion}
\ee
As shown in Ref.~\cite{Goussard:1996dp}, the equation of motion Eq.~\eqref{equationofmotion} remains the same when describing proto-neutron stars at fixed entropies per baryon $s_{B}$. In this case, there is no entropy gradient throughout the star, i.e., $\partial_{i} s_{B}= 0$, where $i$ stands for the spatial coordinates $(r, \theta, \phi)$. Consequently, the additional term $Te^{-H}\partial_{i} s_{B}$ present in the equation of motion (see Eq.4 in Ref.~\cite{Goussard:1996dp}) for hot stars disappears and the standard numerical procedure (as described here) can be used both for cold and hot stars.  Finally, let us notice that the special case with $T=const$ is not realistic, since one expects higher temperatures at higher densities in stars. 

The logarithm of the dimensionless relativistic enthalpy per baryon $H(r,\theta)$ is:
\be
H = \ln \left( \frac{e+p}{m_{b}n_{b}c^{2}  } \right),
\label{enthalpy}
\ee
with the mean baryon mass $m_{b} = 1.66\times10^{-27} \,\rm{kg}$,  and the baryon number density $n_{b}$ . 
At last, the magnetic potential $M(r,\theta)$ in Eq.~\eqref{equationofmotion}, which is associated with the Lorentz force, can be expressed as:
\be
M \left(r, \theta \right) = M \left( A_{\phi} \left(r, \theta \right) \right): = - \int^{0}_{A_{\phi}\left(r, \theta \right)} f\left(x\right) \mathrm{d}x,
\ee
with a current function $f(x)$ as defined in Ref.~\cite{Bonazzola:1993zz} (see Eq. (5.29)). The magnetic star models are obtained by assuming a constant value for the dimensionless current functions $f_{0}$. For higher values of the current function $f_{0}$, the magnetic field in the star increases proportionally. In addition, $f_{0}$ is related to the macroscopic electric current through the relation $j^{\phi}= \Omega A_{t} + (e+p)f_{0}$. According to Ref.~\cite{Bocquet:1995je}, other choices for $f(x)$ different from a constant value are possible, but the general conclusions remain the same.

\section{Results}
In order to model stationary and axi-symmetric neutron and proto-neutron stars in presence of strong poloidal magnetic fields, we solve the coupled Einstein-Maxwell field equations by using the equations of state shown in Fig.~\ref{eos}.
As we are interested in studying how the internal properties of isolated proto-neutron stars change over time, we have fixed the stellar baryonic mass to be $M_{B}=2.35\,M_{\odot}$. This value of $M_{B}$ represents a star whose gravitational mass is close to the maximum mass allowed by TOV solutions of neutron and proto-neutron stars described within this model.  At a fixed baryon mass, one can compare how strangeness (through the presence of hyperons) and neutrinos are distributed inside the star according to the star's temporal evolution. 

The magnetic equilibrium configurations are determined by the choice of the current function $f_{0}$. In table \ref{tabela}, we show the corresponding central baryon number density and the central magnetic field reached in  stars of a given $f_{0}$. Increasing the value of $f_{0}$ arbitrarily, we will find a point where the magnetic force will push the matter off-center so strongly that a topological change to a toroidal configuration take place \cite{cardall2001effects}. As our current numerical tools do not enable us to solve such equilibrium configurations, there is a limit for the magnetic field strength that one can obtain within this approach. In this work, we obtain a maximum current function close to $f_{0}=39000$, which corresponds to a central magnetic field $\sim$ $10^{18}$ G (see table \ref{tabela}) in all three approximate stages of evolution.

\begin{figure}
\begin{center}\includegraphics[width=0.7\textwidth,angle=-90,scale=0.5]{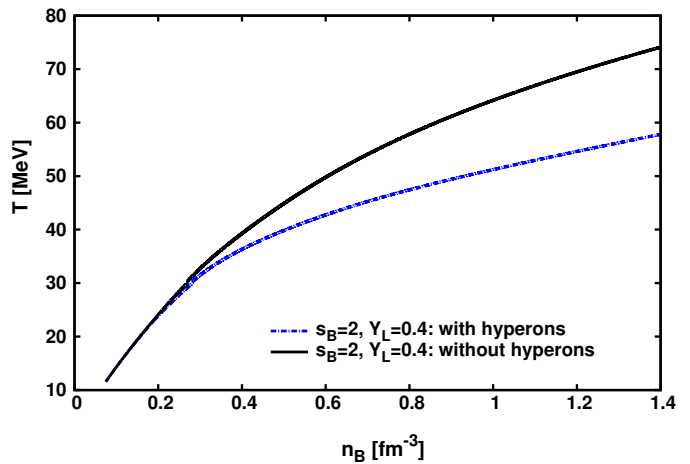}
\caption{\label{eos_hyperons} Temperature as a function of baryon number density for proto-neutron stars with and without hyperons in the EOS.}  
\end{center}
\end {figure}

\begin{table}[h!]
\caption{Relation between the current function $f_{0}$, central baryon number density $n_{B}^c$, the central $B_{c}$ and the surface magnetic field $B_{s}$, and the gravitaional mass $M_{g}$ for a star at fixed baryon mass of $M_{B}=2.35\,M_{\odot}$. We considered three different approximate evolution states  to a hot proto-neutron star from a cold neutron star.} 
\begin{center} \label{tabela}
\begin{tabular}{|c|c|c|c|c|c|} 
\hline
\textbf{EoS}&\textbf{$f_{0}$} & \textbf{$n_{B}^{\,c}(fm^{-3})$} & \textbf{ $B_{c} (10^{18}G)$} & \textbf{ $B_{s} (10^{18}G)$} & \textbf{ $M_{g} (M_{\odot})$}  \\ 
\hline 
     			    &  0     &  0.694  &   0 &  0 & 2.03\\
T=0  		     	&  35000 &  0.509  &  1.01& 0.36& 2.07  \\
	      	        &  39000 &  0.424  &  1.07 & 0.46& 2.09\\
\hline
		      	    &  0   &   0.721  &  0 &  0 &  2.04\\
$s_{B}$=1  &  35000 &  0.514 & 1.02& 0.34& 2.08   \\ 
 $\beta$ 		    &  39000 & 0.402   & 1.06&  0.45& 2.11 \\
\hline
				         &  0     & 0.790 & 0& 0& 2.01\\
$s_{B}$=2  &  35000 &  0.575 & 1.04&0.37& 2.04\\
$Y_{L}$=0.4  				     &  39000 &    0.474 & 1.10& 0.47& 2.06\\
      \hline
\end{tabular}
\end{center}
\end{table}
Throughout this work, we make use of equations of state with hyperon degrees of freedom.
Hyperons are usually not stable and decay into nucleons through the weak interaction in vacuum. However, the condition of $\beta$-equilibrium naturally leads to the existence of hyperons in compact stars, as their decay is Pauli-blocked 
 \cite{ambartsumyan1960degenerate, bethe1974dense, ellis1995kaon,schaffner1996hyperon, prakash1997composition}. The maximum gravitational mass of cold beta-equilibrated matter TOV solution with hyperons is $2.08\,M_{\odot}$ (the corresponding baryon mass is $M_{B}=2.41\,M_{\odot}$), while without hyperons the  gravitational mass reaches $2.14\,M_{\odot}$ (the corresponding baryon mass is $M_{B}=2.50\,M_{\odot}$). In order to investigate the effects of hyperons in proto-neutron stars, in Fig.~\ref{eos_hyperons} we show  equations of state with and  without hyperons for the hadronic chiral SU(3) model with $s_{B}=2$, $Y_{L}=0.4$.  From Fig.~\ref{eos_hyperons}, as the baryon number density increases, the EoS with hyperons becomes colder than the nucleonic one, as already pointed out in Ref.~\cite{Oertel:2016xsn}. This is an effect of the increased number of degrees of freedom and the softening of the EOS with hyperons. 

\begin{figure}
\begin{center}\includegraphics[width=0.7\textwidth,angle=-90,scale=0.5]{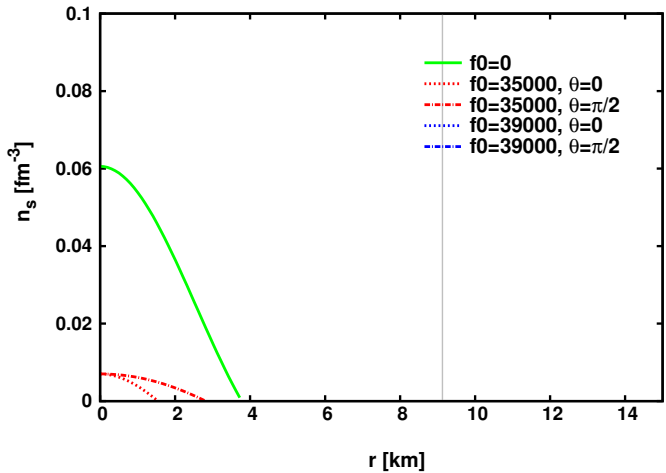}
\caption{\label{ns_t0} Strangeness density profile on equatorial ($\theta=\pi/2$) and polar coordinate radii ($\theta=0$) for one neutron star (T=0 in $\beta$-equilibrium) at fixed baryon mass of $M_{B}=2.35\,M_{\odot}$ (see Table \ref{tabela} for the corresponding gravitational masses). Different $f_{0}$'s correspond to different current functions and characterize different magnetic field profiles. For the largest $f_{0}$ values, the stars possess no strangeness.  }  
\end{center}
\end {figure}

Hyperons contain one or more strange quarks as their internal constituents. This enables us to study how strangeness is distributed inside stars. For example, in Fig.~\ref{ns_t0}, we depict the strangeness density $n_{s}$, which is defined as the sum over the amount of strangeness of each baryon species multiplied by its number density,  as a function of the stellar coordinate radius $r$ for a cold neutron star (T=0 in $\beta$-equilibrium) and at fixed baryon mass of $M_{B}=2.35\,M_{\odot}$. In this figure, the vertical line represents the stellar surface with the corresponding equatorial coordinate radius of $r_{eq}=9.13$ km. The circular coordinates radius $R_{circ}$ represents the equatorial star radius as measured by an observer at infinity and is defined as:
\be
R_{circ} = B(r_{eq},\frac{\pi}{2}) r_{eq},
\ee
with $B(r, \theta)$ being a metric potential (see e.g. Ref.~\citep{Bonazzola:1993zz} for more details). For the star in Fig.~\ref{ns_t0}, one has $R_{circ}=12.37$ km. We have chosen to show all quantities as function of the coordinate radius since there is no appropriate definition for the circular coordinate radius in the polar direction. 

Hyperons are supposed to appear inside  cold, beta-equilibrated, neutrino-free stellar matter at a density of about 2 times nuclear saturation density. According to Fig.~\ref{ns_t0}, the magnetic field changes significantly the amount of strange matter in neutron stars. In particular, strangeness disappears completely for a central magnetic field strength of $\sim 10^{18}$ G (see table \ref{tabela}).  The Lorentz force acts outwards and reduces the stellar central baryon density, so that its value is below the threshold for the creation of hyperons, which are, therefore, suppressed inside the star.

\begin{figure}
\begin{center}\includegraphics[width=0.7\textwidth,angle=-90,scale=0.5]{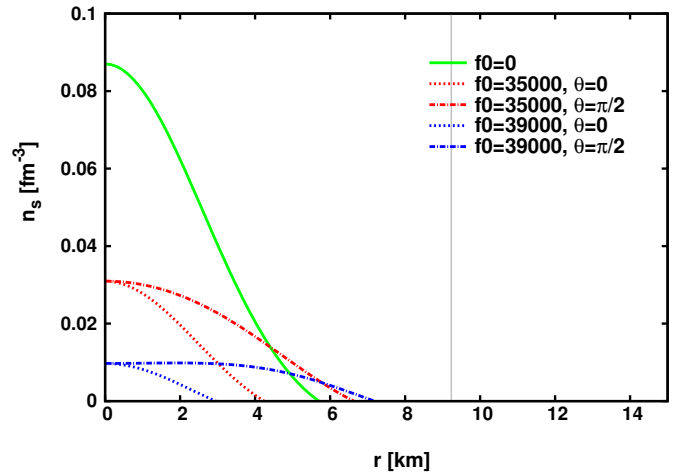}
\caption{\label{ns_s1_beta} Same as in Fig.~\ref{ns_t0} but for one proto-neutron star with fixed entropy per baryon of $s_{B}=1$ in $\beta$-equilibrium. In this case, $M_{B}=2.35\,M_{\odot}$ (see Table \ref{tabela} for the corresponding gravitational masses).}  
\end{center}
\end {figure}

\begin{figure}
\begin{center}\includegraphics[width=0.7\textwidth,angle=-90,scale=0.5]{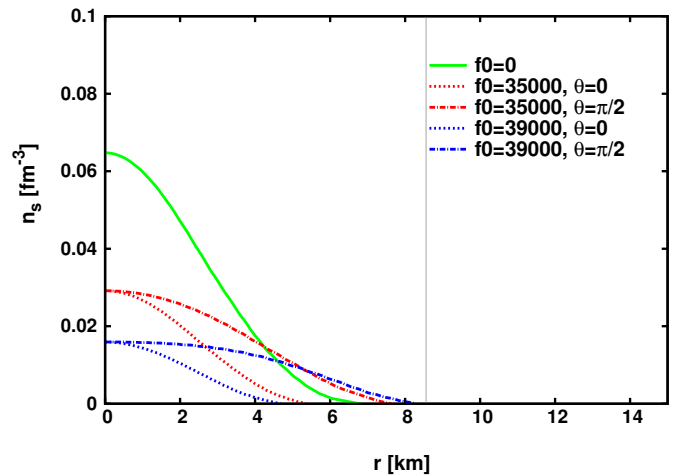}
\caption{\label{ns_s2_yl0.4} Same as in Fig.~\ref{ns_s1_beta} but for one proto-neutron stars with $s_{B}=2$ and fixed lepton fraction $Y_{L}=0.4$. In this case, $M_{B}=2.35 \,M_{\odot}$ (see Table \ref{tabela} for the corresponding gravitational masses).}  
\end{center}
\end {figure}

In Fig.~\ref{ns_s1_beta} and Fig.~\ref{ns_s2_yl0.4} we depict the strangeness density profile as a function of the coordinate radius for proto-neutron star matter in two situations: hot with $s_{B}=1$ and in $\beta$-equilibrium and at very high entropy per baryon $s_{B}=2$ with trapped neutrinos $Y_{L}=0.4$.  
In Fig.~\ref{ns_t0}, Fig.~\ref{ns_s1_beta} and Fig.~\ref{ns_s2_yl0.4}, we show the strangeness density profile on equatorial ($\theta=\pi/2$) and polar directions ($\theta=0$).  For spherical stars, the amount of strangeness is the same in all directions. However,  since the magnetic field breaks the spherical symmetry, magnetized stars will be deformed with respect to the symmetry axis. In this case, they will become oblate objects with polar radius  ($\theta=0$) smaller and equatorial radius ($\theta=\pi/2$)  which will be larger than in the case without magnetic fields. As a result, strangeness will be asymmetrically distributed throughout the star.  For higher values of the magnetic field, the strangeness density can be considered almost constant for a large range of radii, see e.g. Fig.~\ref{ns_s1_beta} for a central field of $\sim 10^{18}$ G ($f_{0}$=39000) and $\theta=\pi/2$.  

The presence of hyperons in neutron or proto-neutron stars may change the neutron star cooling rates \cite{prakash1992rapid, pethick1992cooling, Chatterjee:2015pua}. Moreover, hyperons may also couple to a superfluid state in high density matter \cite{balberg1999roles}. Since the strangeness is directly related to the amount of hyperons and the corresponding channels for neutrino emission, it will affect the cooling behaviour of the star due to the magnetically induced deformation of the star. A related conclusion was already pointed out in Ref.~\cite{Dexheimer:2011pz} for a spherical star.

As stated in Refs.~\cite{cardall2001effects,Franzon:2015gda}, the Lorentz force can reverse its direction in the equatorial plane in magnetized stars.  The Lorentz force is obtained from the derivative of the magnetic potential $M(r, \theta)$ (see Fig.3 in Ref.~\cite{Franzon:2015gda}), which has a minimum at some radius inside the star. This means that the Lorentz force will chance its sign inside the star and, therefore, act differently in different parts of the star.  In addition,  if we suppose that the magnetic field decays over time during the magnetic field evolution in proto-neutron stars, we see from  Fig.~\ref{ns_s1_beta} and Fig.~\ref{ns_s2_yl0.4} that the amount of strangeness becomes higher in the inner core of the star, but it is reduced in the outer layers (crossing lines).  This behaviour is not seen for cold neutron stars, where the strangeness increases in all directions as the magnetic field decays (see Fig.~\ref{ns_t0}).

\begin{figure}
\begin{center}\includegraphics[width=0.7\textwidth,angle=-90,scale=0.5]{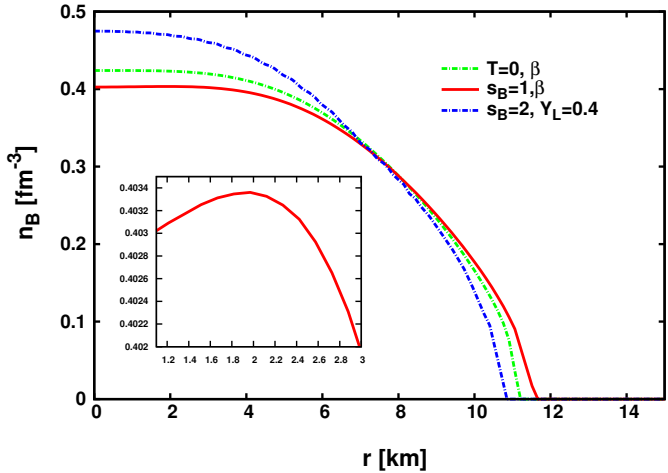}
\caption{\label{dens_profile} Baryon number density profile in the equatorial plane ($\theta=\pi/2$) for the most magnetized  stars studied at fixed baryon mass of  $M_{B}=2.35\,M_{\odot}$ (see Table \ref{tabela} for more details), assuming  3 different approximate stages of evolution: T=0 and $\beta$-equilibrium;  $s_{B}=1$ and $\beta$-equilibrium and  $s_{B}=2$ with $Y_{L}=0.4$. In all cases, the current function is $f_{0}=39000$.}  
\end{center}
\end {figure}

Note that,  for the most magnetized stars studied here (see larger $f_{0}$ in table \ref{tabela}), the maximum density can be reached away from the stellar center. In Fig.~\ref{dens_profile}, we show the baryon number density profile in the equatorial plane for a star with $M_{B}=2.35\,M_{\odot}$ assuming 3 different approximate evolution states: 1) T=0 and $\beta$-equilibrium;  2) $s_{B}=1$ and $\beta$-equilibrium and  3) $s_{B}=2$ with $Y_{L}=0.4$. In the second case, the maximum baryon number density is not at the stellar center.  The others cases do not present this behaviour. This is because stars with lower densities in the inner core become easily more deformed due magnetic fields. Note that the numerical technique presented in Refs.~\cite{Bonazzola:1993zz, Bocquet:1995je} does not handle toroidal configurations as the one in Ref.~\cite{cardall2001effects}. This represents a limit of magnetic field strength that we can obtain within this approach. As one can see from Fig.~\ref{dens_profile}, the maximum baryon number density is shifted away from the center (for the second case), however, this tiny effect is not enough to change the particle population inside stars. Nevertheless, a more comprehensive study  of the subject  would be very desirable by using the formalism from Ref.~\cite{cardall2001effects}.

Neutrinos are mainly produced by electron capture as the progenitor star collapses. However, most of them are temporarily prevented from escaping because their mean free paths are considerably smaller than the radius of the star. This is the well-known trapped-neutrino era, where the entropy per baryon is about 1-2 through most of the star and the total number of leptons per baryon $Y_{L}$ $\backsimeq$ 0.4. As before, we consider neutrino-free and trapped neutrino equations of state either at zero temperature or fixed entropies. In order to model PNS's in their hottest state, we have used $s_{B}=2$ at fixed lepton fraction of $Y_{L}=0.4$. In this case, the neutrinos are trapped and do not leave the star. The amount of neutrinos, on the other hand, depends on the EOS used \cite{Tan:2016ypx}. 

\begin{figure}
\begin{center}\includegraphics[width=0.7\textwidth,angle=-90,scale=0.5]{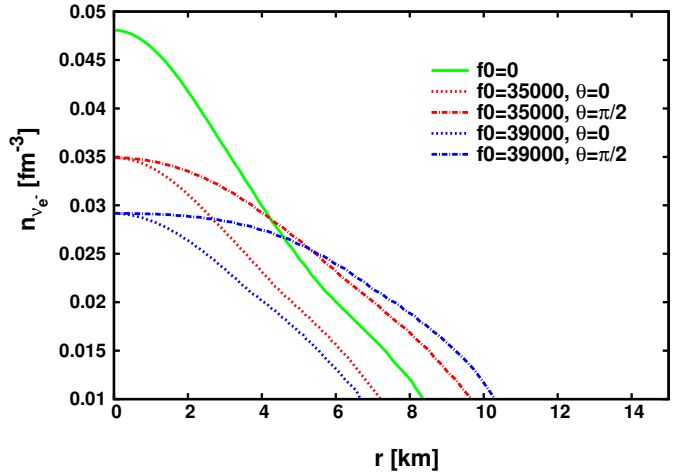}
\caption{\label{neutrino_s2_yl0.4} Electron neutrino density profile as a function of the equatorial and polar coordinate radii for one proto-neutron star with $M_{B}=2.35\,M_{\odot}$ (see Table \ref{tabela} for the corresponding gravitational masses) assuming $s_{B}=2$ and $Y_{L}=0.4$.}  
\end{center}
\end {figure}

In Fig.~\ref{neutrino_s2_yl0.4}, we show the electron neutrino number density profile as a function of the coordinate radius for a star at fixed baryon mass of $M_{B}=2.35 \,M_{\odot}$. This same star is depicted in Fig.~\ref{ns_s2_yl0.4}. The magnetic field reduces the amount of neutrinos present at the center of the star. For example, for the free magnetic field solution, the maximum electron neutrino density is $\sim$ 0.048 $\rm{fm^{-3}}$ at the center of the star. In the maximally magnetized case, this values reduced to $\sim$ 0.030 $\rm{fm^{-3}}$. 
Note that, according to Fig.\ref{neutrino_s2_yl0.4}, the amount of trapped neutrinos decreases as the magnetic field significantly drops for coordinate radii $\gtrsim$ 5 km (in the equatorial plane, $\theta=\pi/2$). However, the opposite effect is seen for radii $\lesssim$ 5 km.  
In addition, since the stars are deformed due to the magnetic field, they become oblate, with a polar radius smaller than the equatorial one. As a consequence, the neutrino flux leaving the PNS's will be asymmetric, having different values in the polar and equatorial directions. These differences may have an observable impact on the neutrino flux from magnetized PNS's. This will be addressed in a future publication. 

\begin{figure}
\begin{center}\includegraphics[width=0.7\textwidth,angle=-90,scale=0.5]{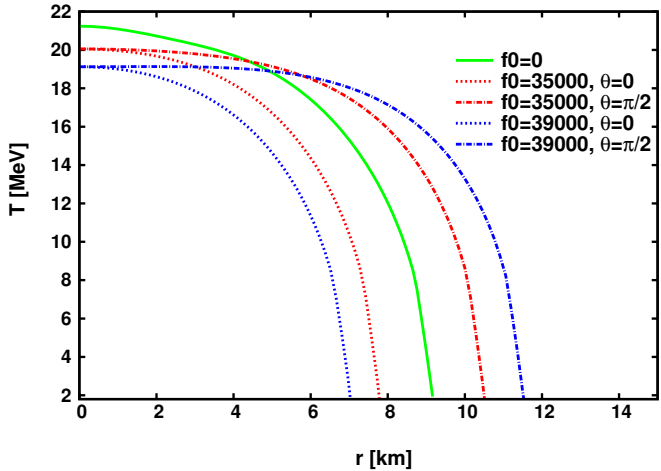}
\caption{\label{temp_s1_beta} Temperature profile as a function of the equatorial and polar coordinate radii for one proto-neutron star with $M_{B}=2.35\,M_{\odot}$ (see Table \ref{tabela} for the corresponding gravitational masses) assuming $s_{B}=1$ and $\beta$-equilibrium.}  
\end{center}
\end {figure}

\begin{figure}
\begin{center}\includegraphics[width=0.7\textwidth,angle=-90,scale=0.5]{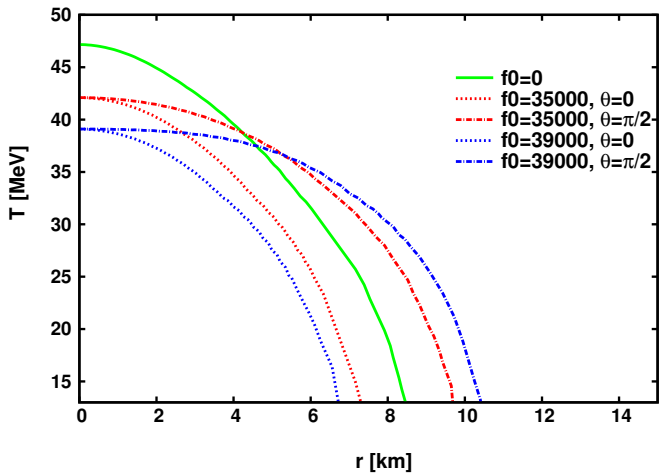}
\caption{\label{temp_s2_yl0.4} Same as in Fig.~\ref{temp_s1_beta} but for one proto-neutron star with $s_{B}=2$ and $Y_{L}=0.4$. In this case, $M_{B}=2.35\,M_{\odot}$ (see Table \ref{tabela} for the corresponding gravitational masses).}
\end{center}
\end {figure}

In Figs.~\ref{temp_s1_beta} and \ref{temp_s2_yl0.4}, we show the temperature throughout a PNS for two approximate stages that
reproduce significant temporal evolution stages. In Fig.~\ref{temp_s2_yl0.4}, the expected initial star (just after the bounce) is lepton rich and extremely hot. For a non-magnetized and spherical star, the temperature at the center reaches values close to 50 MeV (see Fig.~\ref{temp_s2_yl0.4}). On the other hand, when the strong magnetic field is included, the central temperature reaches values below 40 MeV. This same effect is observed (with lower values) for a hot and $\beta$-equilibrated PNS model (see Fig.~\ref{temp_s1_beta}).   

In Fig.~\ref{temp_s1_beta}, the difference in the central temperature between the non-magnetized and the highest magnetized solution is of the order of 2 MeV, much less than in the neutrino $\nu$ trapped era. This is related to the stiffness of the equation of state. According to our model, the equation of state describing the first  approximate stage of evolution is softer than in the other stages. 

According to Refs.~\cite{prakash2001evolution, reddy1998neutrinos, Pons:2000xf, prakash2001evolution}, larger lepton fraction $Y_{L}$ disfavours hyperonic degree of freedom in the stellar interior. As a result, the respective EOS's becomes stiffer. This can be seen in Ref.~\cite{prakash2001evolution}, where a lot of hyperons were present in a $\beta$-equilibrated matter. In our approach, the couplings do not favor a large amount of hyperons in $\beta$-equilibrated matter. In this case, the main effect of fixing  $Y_{L}$ is to make the star more isospin symmetric and, as a consequence of a softer EOS, less massive.

For a PNS with $s_{B}=2$ and $Y_{L}=0.4$, the surface temperature of the core is $\sim$ 13 MeV while for $s_{B}$=1 and $\beta$-equilibrium it is $\sim$ 2 MeV. However, in both cases,  with the decay of  the magnetic field, the temperature increases in the inner layers of the star and decreases in the outer layers. Note that, the same effect was observed for the strangeness density and neutrino distribution inside the star.   

The presence of strong  magnetic fields affects the star surface thermal distribution (see for example Ref.~\cite{aguilera20082d}). The knowledge of the correct temperature distribution in PNS's and NS's is crucial for modelling the cooling of these stars.  Thus, models that include the presence of high magnetic fields should be reconsidered, not only to investigate the effects of the anisotropy of the energy-momentum tensor due to the magnetic field, but also to include the asymmetric  temperature distribution in these objects.  

In addition to magnetic fields, rotation can contribute to the breaking of the spherical symmetry.  In the cases studied here, we have seen that the magnetic field not only affects the macroscopic structure of stars, but also it impacts their microscopic compositions. Such a study is extremely important if one wants to understand the thermal evolution of stellar systems where spherical symmetry is broken.   

The structure of rotating stars is much more complicated than the structure of their non-rotating counterparts \cite{weber1999pulsars,komatsu1989rapidly,cook1994rapidly,stergioulas1994comparing,Bonazzola:1993zz}. The complication comes from a flattening at the poles with an increase of the radius in the equatorial plane. As in the magnetized case, this deformation leads to a dependence of the star's metric both on the polar coordinate $\theta$ and the radial coordinate $r$. 

Although PNS's are probably strongly differentially rotating \cite{baumgarte1999maximum, morrison2004effect,ansorg2009solution}, we model uniformly rotating stars in order to estimate the effect of rotation on strongly magnetized stellar models within a fully general relativity calculation. The effect of the centrifugal force due to rotation in neutron stars was considered already by many authors, see e.g. Refs.~\cite{cook1994rapidly, nozawa1998construction, salgado1994high, negreiros2010modeling, komatsu1989rapidly, weber1991structure} . However, only few works presented self-consistent calculations taking into account both  magnetic field and rotation effects on the neutron star structure \cite{Bocquet:1995je, pili2014axisymmetric, frieben2012equilibrium}.

\begin{figure}
\begin{center}\includegraphics[width=0.7\textwidth,angle=-90,scale=0.5]{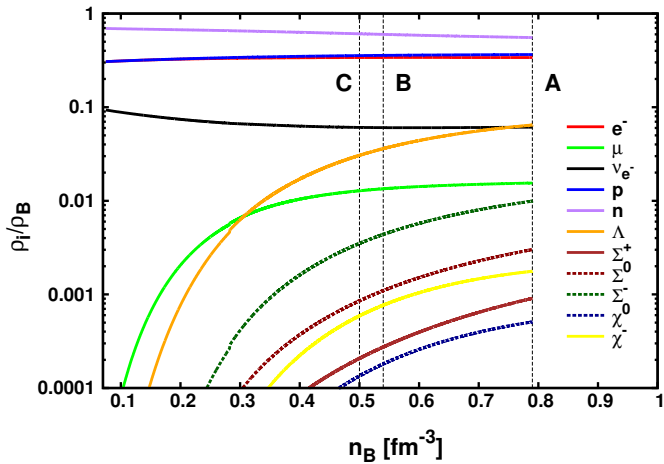}
\caption{\label{pop} Particle population obtained for one proto-neutron star in 3 different cases: $A)$ non-rotating and non-magnetized; $B)$ rotating  at a frequency of 900 Hz and non-magnetized (the corresponding gravitational mass is $M_g=2.04\,M_{\odot}$) and $C)$ rotating and magnetized (the corresponding gravitational mass is $M_g=2.05\,M_{\odot}$). In all cases, the baryonic mass is fixed at $M_B=2.35\,M_{\odot}$ assuming $s_{B}=2$ and $Y_{L}=0.4$.}  
\end{center}
\end {figure}

In Fig.~\ref{pop}, we show the internal composition of proto-neutron stars in 3 scenarios: $A)$ a non-rotating and non-magnetized proto-neutron star at fixed baryon mass of $M_B=2.35\,M_{\odot}$. Matter is described by the EOS with $s_{B}=2$ and trapped neutrinos $Y_{L}=0.4$; $B)$  the same star as in $A)$, but rotating at a frequency of  900 Hz. This frequency is used since the star becomes strongly deformed and it allows us to better study the effects of rotation on the microscopic properties of proto-neutron stars. The results of this analysis can be generalized to other frequencies. Finally, we include  the magnetic field  in the solution $B)$. In this case, we obtain a rotating and magnetized proto-neutron star model denoted by $C)$ for the maximum value of the magnetic field achieved with the code. In this star,  the maximum central magnetic field is 3.76$\times10^{17}$ G. Note that this maximum magnetic field lies below the value obtained for non-rotating proto-neutron stars $\sim 10^{18}$ G for the same baryon mass. In Fig.~\ref{pop}, the particles on the left side of the dashed black lines $A$, $B$ and $C$ represent the populated degrees of freedom inside the corresponding PNS.   

The centrifugal force due to rotation  pushes the matter outwards. As a consequence, the star expands in the equatorial direction and decreases the central number density. For example, in the case $A)$ the baryon  density at the center is $0.790\,\rm{fm^{-3}}$. But if this star rotates at 900 Hz (case $B)$, one obtains a central density of $0.541\,\rm{fm^{-3}}$. And, finally, the corresponding rotating and magnetized star $C)$ yields a central baryon number density of $0.497\,\,\rm{fm^{-3}}$. For stronger magnetic fields such an effect is further increased.  

From Fig.~\ref{pop}, we see that the amount of electron neutrinos is not reduced in rotating PNS's. On the other hand, exotic particles are almost suppressed inside these objects. They might vanish completely in stars rotating faster than the case considered here. Moreover, the magnetic field (in case $C$) further reduces the central number density and, therefore, further modifies  the internal degrees of freedoms. 

\begin{figure}
\begin{center}\includegraphics[width=0.8\textwidth,angle=-90,scale=0.6]{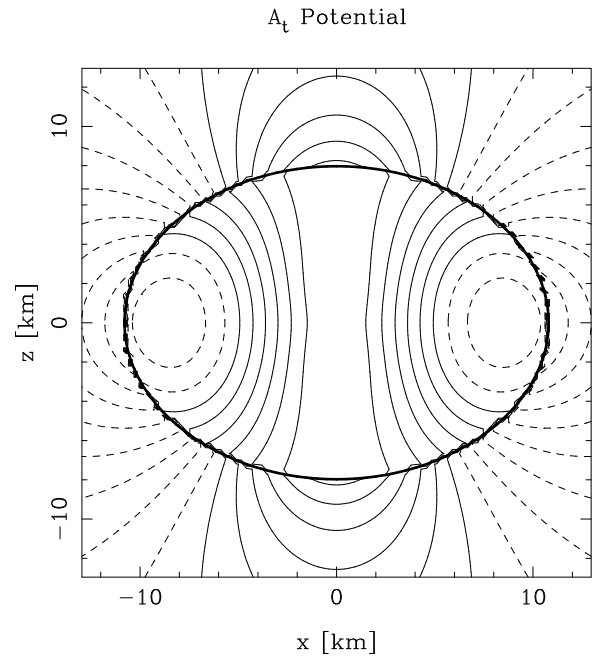}
\caption{\label{rotation_nb} Isocontours of electromagnetic potential lines for a star at fixed $M_B=2.35\,M_{\odot}$ assuming $s_{B}=2$ and $Y_{L}=0.4$ (star $C)$ in Fig.~\ref{pop}). This star rotates at a frequency of 900 Hz and has a central magnetic field of 3.76$\times 10^{17}$ G (the corresponding surface magnetic field is  1.37$\times 10^{17}$ G ) . The ratio between the polar and equatorial  radii is $r_{p}/r_{eq}=0.74$.}   
\end{center}
\end {figure}

In order to visualize the magnetic field distribution, in Fig.~\ref{rotation_nb} we show the electromagnetic potential lines $A_{t}$ isocontours in the $(x, z)$ plane for the star $C)$.  This star corresponds to a gravitational mass of $M_{g}=2.05\,M_{\odot}$.  The centrifugal and magnetic forces act against gravity, which allows stars to be more massive than their non-magnetized or non-rotating counterparts. 

\section{Conclusion}
We have computed models of  massive and highly magnetized neutron and proto-neutron stars  in a fully general relativistic framework. Under the standard assumption that PNS's undergo a more quiet and quasi-stationary evolution after their birth,  we investigated the role played by magnetic fields and rotation on the surface deformation and on their internal matter distribution. This study represents the first step towards a fully self-consistent treatment of the cooling of neutron stars that have their spherical symmetry broken due to strong magnetic fields. 

In order to do so, we solved the Einstein-Maxwell equations self-consistently for stars at fixed baryon mass of $M_{B}=2.35\,M_{\odot}$. We have chosen to do so because this value represents a stellar mass close to the maximum mass for non-magnetized and spherical configurations. We included poloidal magnetic fields which are generated by macroscopic currents taking into account the anisotropies associated with such a field. In this study, we have not included the effects of the magnetic field on the equation of state through the effects of Landau quantization of the charged particles in the presence of a magnetic field, since it was already found that this contribution to the macroscopic properties of stars is small compared to the pure field contribution of the energy-momentum tensor \citep{Chatterjee:2014qsa, franzon2016self}.
 
We investigated the equation of state of cold NS  and warm PNS matter in the neutrino-free and neutrino-trapped scenarios making use of the  hadronic chiral SU(3) model. We then determined the properties of PNS's and NS's consisting of hadronic matter with  hyperons. The calculations were performed  for zero temperature and at fixed entropy per baryon. Our results indicate that spherical hot stars with trapped neutrinos, i.e, $s_{B}=2$ and $Y_{L}=0.4$, are less massive than the same stars in $\beta$-equilibrium or their cold counterparts.

The primary effect of the magnetic field decay is to increase the amount of neutrinos and the strangeness at the stellar core. As the magnetic field decreases, we see also an increase of the temperature at the stellar center.  Note that, assuming that the magnetic field decays over time, the temperature in the equatorial plane increases in the inner core while it decreases in the outer core. This fact is related to the Lorenz force, which reverses its direction in the equatorial plane.  

As shown in Refs.~\cite{markey1973adiabatic, tayler1973adiabatic, wright1973pinch}, magnetic field geometries with purely poloidal or purely toroidal magnetic field configurations undergo the  so-called  Tayler instability. Recently,  such instabilities were confirmed both in Newtonian numerical simulations \cite{lander2012there, braithwaite2006stability, braithwaite2006stable, braithwaite2007stability} and in general relativity framework in Refs.~\cite{ciolfi2013twisted, lasky2011hydromagnetic, marchant2011revisiting,  mitchell2015instability}.  According to Refs.~\cite{armaza2015magnetic, prendergast1956equilibrium, braithwaite2004fossil, braithwaite2006stable, akgun2013stability}, equilibrium magnetic field configurations are possible for a twisted-torus 
geometry, with  poloidal and toroidal magnetic field components. It is to be noted, that the magnetic flux might change its strength and, therefore, its distribution in the star due to dissipation of the electric current   \cite{goldreich1992magnetic}.  Although we have assumed a purely poloidal magnetic field in this work,  we can have a fair idea of the maximum magnetic field strength that can be reached inside these objects and also understand the effects of strong magnetic field on the microphysics of PNS's.

We further studied the properties of PNS's subjected to fast rotation. Our results indicate that the electron neutrino distribution of rotating proto-neutron stars does not differ  much 
from their non-rotating counterpart.  This is possible due to the fact that the centrifugal forces  ($f_{c} \propto r\Omega^2$) act mainly on the outer layers of the star. However, the amount of hyperons is reduced inside these objects, what may affect the cooling of these stars.  We  have also included magnetic fields in the rotating PNS  model. As expected, the reduction in the central densities is even more pronounced and magnetic fields suppress exotic phases in rotating PNS even further, as in the case of cold neutron stars. A combination of both magnetic field and rotation effects can impact, for example, the nucleosynthesis of the winds in PNS's \citep{vlasov2014neutrino}.

In addition, the scenario of transformation of a proto-neutron star into a neutron star could be influenced by a quark-hadron phase transition due to the presence of high temperatures, in which case the transition happens at lower densities. Such
stars would be composed of hot quark and hadronic matter at different leptons fractions and fixed entropies. It would be interesting to couple our results to a hybrid star scenario with a quark-hadron phase transition in the star core.  Moreover, the cooling behaviour strongly depends on the particle composition of the star, which determines  neutrino emission channels.  This will also be considered in a future work together with effects related to the proto-neutron star crust.

\begin{acknowledgements}
 B. Franzon acknowledges support from CNPq/Brazil, DAAD and HGS-HIRe for FAIR.  B. Franzon gratefully acknowledges the kind hospitality at the Department of Physics of the Kent State University, and support through  HGS-HIRe for FAIR Fellowship for Research Abroad. S. Schramm acknowledges support from the HIC for FAIR LOEWE program. The authors wish to acknowledge the "NewCompStar" COST Action MP1304.
\end{acknowledgements}

\nocite{*}

\bibstyle{apsrev4-1}
\bibliography{pns_biblio}

\end{document}